\documentclass[10pt,aps,twocolumn,nobibnotes,pra,showpacs]{revtex4-1}
  \usepackage[dvips]{color}
  \usepackage{newlfont}
  \usepackage{graphicx}
  \usepackage{amssymb}
  \usepackage{amsmath}
  \usepackage[latin1]{inputenc}
	\usepackage{bbm}
	\usepackage{mathptmx}

\newcommand{\be}{\begin{eqnarray}}
\newcommand{\ee}{\end{eqnarray}}
\newcommand{\ket}[1]{\mbox{$ | #1 \rangle $}}
\newcommand{\bra}[1]{\mbox{$ \langle #1 | $}}

\newcommand{\h}{\scriptscriptstyle \mathrm{H}}

\begin{document}

\title{Galilei covariance and Einstein's equivalence principle in quantum reference frames}
\author{S.~T.~Pereira}
\author{R.~M.~Angelo}\email{renato@fisica.ufpr.br}
\affiliation{Department of Physics, Federal University of Paran\'a, P.O. Box 19044, 81531-980, Curitiba, PR, Brazil}

\begin{abstract}
The covariance of the Schr\"odinger equation under Galilei boosts and the compatibility of nonrelativistic quantum mechanics with Einstein's equivalence principle have been constrained for so long to the existence of a superselection rule which would prevent a quantum particle from being found in superposition states of different masses. In an effort to avoid this expedient, and thus allow nonrelativistic quantum mechanics to account for unstable particles, recent works have suggested that the usual Galilean transformations are inconsistent with the nonrelativistic limit implied by the Lorentz transformation. Here we approach the issue in a fundamentally different way. Using a formalism of unitary transformations and employing quantum reference frames rather than immaterial coordinate systems, we show that the Schr\"odinger equation, although form variant, is fully compatible with the aforementioned principles of relativity.
\end{abstract}

%03.65.Ta	Foundations of quantum mechanics; measurement theory
%02.20.Uw	Quantum groups
%11.30.-j	Symmetry and conservation laws
%\keywords{Quantum reference frames, Galilean covariance, Einstein's equivalence principle}

\maketitle

%--------------------------------------------------------
\section{Introduction\label{introduction}}
\vspace{-0.2cm}

In 1954, Bargmann published a paper~\cite{bargmann54} showing that the nonrelativistic quantum mechanics, as described by the Schr\"odinger equation, is {\em not} trivially covariant under Galilei boosts. According to him, {\em covariance}---here understood as {\em form invariance}---can be attained only by attaching a mass-dependent phase to the wave function. This procedure, however, yields a dramatic collateral effect: in order to guarantee covariance upon a cyclic sequence of Galilean transformations, which brings the description back to the original reference system, a superselection rule must act in order to prevent the existence of mass superpositions. From the same arguments, it follows that quantum mechanics cannot coexist peacefully with Einstein's equivalence principle unless such a superselection rule exists. The essence of Bargmann's argument can be put as follows (see Refs.~\cite{greenberger01,okon13} for recent treatments).

Consider the Schr\"odinger equation for a particle $\mathbb{P}$ moving freely in one dimension from the perspective of an {\em inertial reference system} $\mathbb{S}_0$:
\be
i\hbar \,\partial_t \psi(x,t)=-\frac{\hbar^2}{2m}\,\partial_x^2 \psi(x,t).
\label{SEx}
\ee
The physics from the viewpoint of a distinct reference system $\mathbb{S}$, whose instantaneous position relative to $\mathbb{S}_0$ is $X(t)$, is obtained via the Galilean transformation 
\be
x'=x-X(t) \quad \text{and} \quad t'=t,
\label{x't'}
\ee
where $(x',t')$ are the spacetime coordinates in $\mathbb{S}$. It follows that $\partial_{x}=\partial_{x'}$, $\partial_{t}=\partial_{t'}-\dot{X}\partial_{x'}$, and
\be
i\hbar\,\partial_{t}\psi'(x',t)=\left(-\frac{\hbar^2}{2m}\,\partial_{x'}^2+i\hbar\dot{X}\partial_{x'} \right)\psi'(x',t),
\ee
where $\psi'(x',t)=\psi(x'-X(t),t)$. We use ``$t$'' for all reference frames, as time is absolute in Galilean relativity. Since the form of the resulting Hamiltonian is not equal to the free-particle Hamiltonian appearing in Eq.~\eqref{SEx}, the Schr\"odinger equation is said to be {\em noncovariant} under a Galilean transformation. In order for covariance to be attained, it is presumed that the {\em correct} wave function must be given by
\be\begin{array}{l}
\varphi'=e^{-i\phi(x',t)}\psi', \\ \\
\phi(x',t)=\frac{m\dot{X}x'}{\hbar}+\frac{m}{2\hbar}\int_{0}^{t} dt\,\dot{X}^2,
\label{varphi'}
\end{array}\ee
for in this case one arrives at
\be
i\hbar\,\partial_t\varphi'=\left(-\frac{\hbar^2}{2m}\partial_{x'}+m\ddot{X}\hat{x}'\right)\varphi'.
\label{covariantSE'x}
\ee
Clearly, {\em Galilean covariance} (GC) is retrieved upon boosts, i.e., when $\dot{X}$ is constant. The appearance of a gravitational field in the perspective of an accelerated frame $\mathbb{S}$ is, on the other hand, the expression of {\em Einstein's equivalence principle} (EEP). 

Bargmann exploited the consequences of such a mass-dependent phase using the following rationale. Assume that a quantum particle has been prepared in a superposition of masses,
\be
\psi \propto \psi_{m_1}+\psi_{m_2}.
\label{superposition}
\ee
Consider a cyclic Galilean transformation $G_c$, i.e., a sequence of transformations that leads the description to the perspective of other reference frames and then brings it back to $\mathbb{S}_0$'s perspective. Specifically, $G_c$ is chosen to be composed of two opposite translations and two opposite boosts:
\be
G_c\equiv G_{-vt}\,G_{-a}\,G_{vt}\,G_a=e^{iavm/\hbar}.
\label{Gc}
\ee
$G_X$ implements the transformation \eqref{x't'}, i.e., $G_X(x,t)=(x-X(t),t)$. When applied to space-time coordinates, the cyclic transformation is equivalent to the identity: $G_c(x,t)=(x,t)$. Applied, however, on a wave function $\psi_m$, it gives $G_c\psi_m=e^{iavm/\hbar}\varphi'_m$. It follows that the cyclic transformation of the superposition \eqref{superposition} yields
\be
G_c\psi\propto e^{iav m_1/\hbar}\varphi'_{m_1}+e^{iavm_2/\hbar}\varphi'_{m_2},
\label{m1m2}
\ee
which has a measurable relative phase $va\Delta m/\hbar$, where $\Delta m=m_2-m_1$. To eliminate such an undesirable effect, Bargmann postulated the superselection rule $\Delta m=0$, meaning that no mass-superposition state can exist.

The framework devised by Bargmann constrains the compatibility of quantum mechanics with both GC and EEP to the nonexistence of mass superpositions. While Bargmann's formal construction is widely accepted in the literature and already figures as textbook content~\cite{ballentine98,yan03}, his conclusion has been recurrently debated~\cite{greenberger01,okon13,leblond63,greenberger70_1,greenberger70_2,greenberger74_1,greenberger74_2,pascual90,giulini96,holland99,brown99,wick07,pad11,coronado12}. In particular, some authors interpreted the mass-dependent phase as a residue of the twin-paradox effect in nonrelativistic quantum mechanics~\cite{greenberger01,coronado12}, thus attributing deeper physical meaning to it. On the other hand, there are some works defending that quantum mechanics is compatible with EEP~\cite{unnik02,sudarsky03,wawr11}. In consonance with a program initiated long ago~\cite{greenberger70_1,greenberger70_2,greenberger74_1,greenberger74_2}, it was recently suggested that the mass-superselection rule can be avoided if mass is taken as a dynamical variable~\cite{okon13}. Regardless of their conclusions, all of the aforementioned works stick to traditional ingredients, namely, the reference frames are fundamentally classical and the analysis is focused on the wave function. 

It is the aim of this contribution to critically revisit the subject within two complementary frameworks. First, in Secs.~\ref{UT} and \ref{RBA}, by suitably employing passive unitary transformations, we show how quantum mechanics is able to respect GC and EEP without appealing to any sort of superselection rule. Then, in Secs.~\ref{LRF} and \ref{QGB}, we push the subject to the conceptually deeper framework of {\em quantum reference frames}~\cite{aharonov84,spekkens07,angelo11,angelo12}. We shall ask how the quantum mechanical description manifests when the frame of reference is itself a physical system describable by the Schr\"odinger equation. Then, by employing nontrivial unitary transformations and correctly interpreting the corresponding dynamics, we show once again that there is no tension between quantum mechanics and the aforementioned relativity principles.

%-------------------------------------------------------------------
\section{Unitary generators of coordinate transformations \label{UT}}
\vspace{-0.2cm}

It is well known that unitary operators are useful tools for describing changes between reference frames. However, the correct application of these objects is subtle and deserves a close look, which is the main purpose of this section.

%---------------------------
\subsection{Passive picture}
\vspace{-0.2cm}
Consider a generic $N$-particle system evolving in time relatively to an inertial reference system $\mathbb{S}_0$ according to the Schr\"odinger equation
\be
i\hbar\,\partial_t\ket{\psi(t)}=H(\hat{\mathbf{r}})\ket{\psi(t)},
\ee
where $\hat{\mathbf{r}}=(\hat{\mathbf{r}}_1,\cdots,\hat{\mathbf{r}}_N)$ and $\hat{\mathbf{r}}_k=(\hat{x}_k,\hat{p}_k)$, with the latter being the phase-space coordinate of the $k$-th particle relative to $\mathbb{S}_0$. A passive change of the coordinates can be implemented by a unitary transformation $\hat{\mathcal{G}}=\mathcal{G}(\hat{\mathbf{r}},t)$ as follows:
\be
\hat{\mathbf{r}}'\equiv\hat{\mathcal{G}}\,\hat{\mathbf{r}}\,\hat{\mathcal{G}}^{\dag}=\mathbf{f}(\hat{\mathbf{r}}),
\label{r'f}
\ee
where $\hat{\mathbf{r}}'$ is the vector composed of the phase-space coordinates relative to moving reference system, $\mathbb{S}$. Since $\hat{\mathcal{G}}$ is unitary, there holds that $\hat{\mathbf{r}}=\mathbf{f}(\hat{\mathcal{G}}^{\dag}\,\hat{\mathbf{r}}\,\hat{\mathcal{G}})$, from which it follows that $\hat{\mathcal{G}}^{\dag}\,\hat{\mathbf{r}}\,\hat{\mathcal{G}}=\mathbf{f}^{-1}(\hat{\mathbf{r}})$, where $\mathbf{f}^{-1}$ denotes the inverse of $\mathbf{f}$. 

In the passive picture, the physical state $\ket{\psi(t)}$ remains unchanged, so that expectation values are computed as $\bra{\psi(t)}\hat{\mathbf{r}}'\ket{\psi(t)}=\bra{\psi(t)}\mathbf{f}(\hat{\mathbf{r}})\ket{\psi(t)}$. The vector state $\ket{\psi}=\int d^N\mathbf{x}\,\psi(\mathbf{x})\ket{\mathbf{x}}_{\hat{\mathbf{x}}}$, originally written in the eigenbasis $\{\ket{\mathbf{x}}_{\hat{\mathbf{x}}}\}$ of $\hat{\mathbf{x}}=(\hat{x}_1,\cdots,\hat{x}_N)$, can be expressed in the new coordinate system through the mapping
\be
\ket{\mathbf{x}}_{\hat{\mathbf{x}}}=\ket{\mathbf{f}(\mathbf{x})}_{\hat{\mathbf{x}}'}.
\label{map}
\ee
Each position eigenstate in the Hilbert space spanned by the eigenvectors of $\hat{\mathbf{r}}$ maps to a counterpart in the Hilbert space spanned by the eigenvectors of $\hat{\mathbf{r}}'$. As an example, consider the usual transformation from laboratory coordinates $(\hat{x}_1,\hat{x}_2)$ to center-of-mass and relative coordinates $(\hat{x}_{cm},\hat{x}_r)$. In this case, the mapping gives $\ket{x_1}_{\hat{x}_1}\ket{x_2}_{\hat{x}_2}=\ket{\tfrac{m_1x_1+m_2x_2}{m_1+m_2}}_{\hat{x}_{cm}}\ket{x_2-x_1}_{\hat{x}_r}$.

The derivation of the transformed Hamiltonian is better illustrated in the Heisenberg picture. From $\bra{\psi(t)}\hat{\mathbf{r}}'\ket{\psi(t)}=\bra{\psi(0)}\hat{U}^{\dag}\hat{\mathbf{r}}'\hat{U}\ket{\psi(0)}$, we define the Heisenberg operator in the moving frame as $\hat{\mathbf{r}}_{\h}'=\hat{U}^{\dag}\hat{\mathbf{r}}'\hat{U}$, where $\hat{U}=e^{-iH(\hat{\mathbf{r}})t/\hbar}$. Noting that $\hat{\mathbf{r}}'$ may explicitly depend on time, one applies a time derivative to $\hat{\mathbf{r}}'_{\h}$ to obtain
\be
\frac{d\hat{\mathbf{r}}'_{\h}}{dt}=\frac{[\hat{\mathbf{r}}'_{\h},\hat{H}'_{\h}(t)]}{i\hbar},
\ee
where $\hat{H}'_{\h}(t)=\hat{U}^{\dag}\hat{H}'\hat{U}$ and
\be
\hat{H}'=H'(\hat{\mathbf{r}}',t)=H(\hat{\mathbf{r}})+i\hbar\,\mathcal{G}(\hat{\mathbf{r}},t)\,\partial_t\mathcal{G}^{\dag}(\hat{\mathbf{r}},t),
\label{H'passive}
\ee
with $\hat{\mathbf{r}}=\mathbf{f}^{-1}(\hat{\mathbf{r}}')$. This is the transformed Hamiltonian to be used in the Schr\"odinger equation. For time-independent unitary generators, we obtain
\be
\hat{H}'=H(\mathbf{f}^{-1}(\hat{\mathbf{r}}'))=\hat{\mathcal{G}}H(\mathbf{f}^{-1}(\hat{\mathbf{r}}))\hat{\mathcal{G}}^{\dag}=\hat{\mathcal{G}}\hat{\mathcal{G}}^{\dag} H(\hat{\mathbf{r}})   \hat{\mathcal{G}}\hat{\mathcal{G}}^{\dag}.\qquad
\label{H'notime}
\ee
We see that this Hamiltonian is precisely the original one rewritten in terms of the moving system coordinates, which is a result that is in total accordance with the prescription given in Ref.~\cite{aharonov84}.

%--------------------------
\subsection{Active picture}
\vspace{-0.2cm}
In an active change, defined by $\ket{\psi'}=\hat{\mathcal{G}}^{\dag}\ket{\psi}$, the unitary generator acts on the vector state while the operators remain unchanged. This corresponds to conserving the original coordinate system but physically changing the quantum state. The result, though, is equivalent to that of passive changes, as is obvious from $\bra{\psi'}\hat{\mathbf{r}}\ket{\psi'}=\bra{\psi}\hat{\mathbf{r}}'\ket{\psi}$. The transformed Schr\"odinger equation reads $i\hbar\,\partial_t\ket{\psi'}=\hat{H}'(t)\ket{\psi'}$, where
\be
\hat{H}'=H'(\hat{\mathbf{r}},t)=\hat{\mathcal{G}}^{\dag}H(\hat{\mathbf{r}})\hat{\mathcal{G}}+i\hbar\,\big(\partial_t\hat{\mathcal{G}}^{\dag} \big)\hat{\mathcal{G}}.
\label{H'active}
\ee
Consider again the case in which $\hat{\mathcal{G}}$ is time independent. It follows that $\hat{H}'=H(\mathbf{f}^{-1}(\hat{\mathbf{r}}))$, which can be directly compared with Eq.~\eqref{H'notime}. We see that the operator $\hat{\mathbf{r}}$ of an active change assumes the same interpretation of $\hat{\mathbf{r}}'$ of a passive change. Accordingly, it can be directly checked that the Hamiltonian operators \eqref{H'passive} and \eqref{H'active} are connected by $H'(\hat{\mathbf{r}}',t)=\hat{\mathcal{G}} H'(\hat{\mathbf{r}},t)\hat{\mathcal{G}}^{\dag}$.

The techniques discussed in this section directly apply to both quantum and classical references frames, a point that will be illustrated in the next sections.

%----------------------------------------------------
\section{Reassessing Bargmann's analysis \label{RBA}}
\vspace{-0.2cm}
We now revisit Bargmann's scheme using unitary generators. Throughout the paper, we work with the passive picture, which gives a more direct interpretation for the moving system coordinates. This strategy aims at highlighting our understanding that a {\em boost} is a mere theoretical change in the description of the system, from $\mathbb{S}_0$'s perspective to $\mathbb{S}$'s, rather than a real propulsion of the reference frame $\mathbb{S}_0$ (see Refs.~\cite{angelo11,angelo12} for similar approaches). We start with the physics seen from the perspective of an inertial reference system $\mathbb{S}_0$: 
\be
i\hbar\,\partial_t\ket{\psi}=\hat{H}\ket{\psi}, \qquad \hat{H}=H(\hat{p})=\frac{\hat{p}^2}{2m}.
\label{H}
\ee 
Consider the unitary operator $\hat{\mathcal{G}}(t)=e^{-iX(t)\hat{p}/\hbar}$. It moves the description to the perspective of a {\em classical} reference system $\mathbb{S}$, which occupies the instantaneous position $X(t)$ relative to $\mathbb{S}_0$. The resulting coordinates read
\be\begin{array}{l}
\hat{x}'=\hat{\mathcal{G}}\,\hat{x}\,\hat{\mathcal{G}}^{\dag}=\hat{x}-X(t), \\ \\
\hat{p}'=\hat{\mathcal{G}}\,\hat{p}\,\hat{\mathcal{G}}^{\dag}=\hat{p}. 
\label{xp'}
\end{array}\ee
Following the prescriptions of the previous section, we obtain the transformed Schr\"odinger equation,
\be
i\hbar\,\partial_t\ket{\psi}=\hat{H}'\ket{\psi}, \qquad \hat{H}'=\frac{\hat{p}'^2}{2m}-\dot{X}\hat{p}',
\label{SE'}
\ee
Since $\hat{H}'\neq H(\hat{p}')$, the Schr\"odinger equation is not Galilei covariant. Following Bargmann's approach, at this point we should attach a mass-dependent phase to the vector state [see Eq.~\eqref{varphi'}]. However, it is clear that no $c$ number would be able to remove the operator $\dot{X}\hat{p}'$ from the Hamiltonian. Instead, we need another unitary transformation, which we can build inspired by the phase given in Eq.~\eqref{varphi'}. But this is equivalent to another change of coordinates. Accordingly, we can return to the starting point and use a composite transformation,
\be
\hat{\mathcal{G}}_X=e^{-iX\hat{p}/\hbar}e^{im\dot{X}\hat{x}/\hbar}e^{i\theta(X)},
\label{GX}
\ee
where $\theta(X)=\frac{m}{2\hbar}\int_{0}^{t} dt\,\dot{X}^2$. Now one has that
\be
\hat{x}'=\hat{x}-X\quad \text{and} \quad \hat{p}'=\hat{p}-m\dot{X}.
\label{xp_relative}
\ee
Using the formula \eqref{H'passive}, we get the passive Hamiltonian
\be
i\hbar\,\partial_t\ket{\psi}=\hat{H}'\ket{\psi}, \qquad \hat{H}'=\frac{\hat{p}'^2}{2m}+m\ddot{X}\hat{x}'.
\label{SE_relative}
\ee
As in the formulation \eqref{covariantSE'x}, it is clear that the resulting equation obeys GC and EEP. 

At this point, however, someone might object that we have used a mass-dependent generator, a questionable expedient to produce a coordinate transformation. But let us return to Eqs.~\eqref{xp'} and \eqref{SE'}. Although they have been produced by a mass-independent transformation, which clearly yields the position of the particle relative to $\mathbb{S}$, it preserves the momentum relative to $\mathbb{S}_0$. It follows, therefore, that the Hamiltonian in Eq.~\eqref{SE'} is {\em not} properly given in terms of the coordinates accessible to $\mathbb{S}$. On the other hand, we see by Eqs.~\eqref{xp_relative} that $\hat{\mathcal{G}}_X$ leads to both the relative position and the correct relative momentum. As a consequence, the Schr\"odinger equation \eqref{SE_relative} emerges as a better candidate to describe the physics seen from $\mathbb{S}$'s perspective. The bottom line is that the Schr\"odinger equation depends on the Hamiltonian of the system, whose formulation focuses on the canonical pair $(\hat{x},\hat{p})$. Since momentum---a primordially mass-dependent quantity---plays an essential role in the Hamiltonian formalism, it is not possible to obtain a fully relative description by dealing only with positions. Then, in accepting the need to also transform the canonical momentum, we necessarily have to admit the appearance of mass in the transformations.

Now the crucial question arises as to whether $\hat{\mathcal{G}}_X$ will lead to a mass-dependent phase in cyclic transformations. The answer is yes, it will in a sense, but the remnant phase is physically irrelevant. This can be proved as follows. Using standard techniques of operator algebra, one may show that
\be
\hat{\mathcal{G}}_{X_2}\hat{\mathcal{G}}_{X_1}=e^{i\Theta_m(X_1,X_2)}\hat{\mathcal{G}}_{X_1+X_2},
\label{Gx2x1}
\ee
where $\Theta_m(X_1,X_2)=\theta(X_1)+\theta(X_2)-\theta(X_1+X_2)+\frac{mX_1\dot{X}_2}{\hbar}$. The expectation value of an arbitrary operator $\hat{\mathcal{O}}$ measured in the shifted reference frame is 
\be\begin{array}{l}
\bra{\psi}\hat{\mathcal{G}}_{X_2}\hat{\mathcal{G}}_{X_1}\,\hat{\mathcal{O}}\,\,\hat{\mathcal{G}}_{X_1}^{\dag}\hat{\mathcal{G}}_{X_2}^{\dag}\ket{\psi}=\bra{\psi}\hat{\mathcal{G}}_{X_1+X_2}\,\hat{\mathcal{O}}\,\,\hat{\mathcal{G}}_{X_1+X_2}^{\dag}\ket{\psi}. \nonumber
\end{array}\ee
It immediately follows, for any cyclic transformation with $\sum_kX_k=0$, that
\be\begin{array}{l}
\bra{\psi}\Big(\prod\limits_{k}\hat{\mathcal{G}}_{X_k}\Big)\hat{\mathcal{O}}\Big(\prod\limits_k\hat{\mathcal{G}}_{X_k}\Big)^{\dag}\ket{\psi}=\bra{\psi}\hat{\mathcal{O}}\,\ket{\psi},
\end{array}\ee
which has no remnant measurable phase. This result, which holds for every vector state $\ket{\psi}$, shows that the original description will always be retrieved under cyclic transformations. This ought to be so, as the transformation $\hat{\mathcal{G}}_{X_k}$ reflects only a change in the theoretical description; it is not {\em real}. That is, absolutely no physical intervention is implied to the system by a cyclic transformation which, naturally, just leads us back to the original description. The actual issue with Eq.~\eqref{m1m2} derives from considering that mass is an operator, a notion that figures as a tacit assumption in Bargmann's approach and a declared model in Refs.~\cite{coronado12,okon13}, and that mass superpositions exist. Accepting that a boost is a theoretical operation implies that the residual phase in Eq.~\eqref{m1m2} should actually be viewed as an indication that there is something wrong with some premise of Bargmann's argument, presumably with the assumptions that mass is an operator and that mass superpositions exist (see discussion below). If we stick to the standard nonrelativistic quantum mechanics, according to which mass is just a parameter, and adopt the framework proposed above, then no issue arises and we can conclude that quantum mechanics is in full harmony with GC and EEP.

In support of this conclusion, two crucial points have to be underlined. First, quantum mechanics is not just about vectors living in a complex space. To make the link with the real world, we need to ask how observables behave when the system is prepared in a given state. Hence, it may be the case that it is not completely fair to demand, solely from Schr\"odinger's equation, compatibility with relativity principles. In fact, we should critically ask: is the Schr\"odinger equation the ultimate {\em law} of quantum physics from which we ought to demand covariance? Consider the Hamiltonian given in Eq.~\eqref{SE'}. Without solving the Schr\"odinger equation, it is not possible to infer the real relevance of the ``unwanted'' term $\dot{X}\hat{p}'$ for the dynamics. Actually, to make accurate statements about the dynamics relative to $\mathbb{S}$, we need to look at $d^2\langle\hat{x}'\rangle/dt^2$. To this end, the Heisenberg picture reveals itself particularly useful. Write the acceleration as $d^2\langle\hat{x}'\rangle/dt^2=\bra{\psi(0)}\ddot{\hat{x}}'_{\h}\ket{\psi(0)}$. In the referred picture, the motion of $\mathbb{P}$ from $\mathbb{S}_0$'s perspective is governed by the relation $\ddot{\hat{x}}_{\h}=0$, which is derived from the Hamiltonian \eqref{H}. Now using the Hamiltonian operators given in Eqs.~\eqref{SE'} and \eqref{SE_relative}, we obtain from both that
\be
\ddot{\hat{x}}'_{\h}=-\ddot{X},
\ee 
which clearly respect GC for $\dot{X}$ constant and EEP for $\ddot{X}$ constant. Therefore, while the Schr\"odinger equation \eqref{SE'} is not covariant, it is not correct to make the same statement for the equations of motion it generates. This point can be further appreciated via the Hamiltonian
\be
\hat{H}'(\alpha)=\frac{(\hat{p}'-\alpha\,m\dot{X})^2}{2m}+(1-\alpha)\,m\ddot{X}\hat{x}',
\label{generalH'}
\ee
where $\alpha$ is a real number. Heisenberg's equations give $m\dot{\hat{x}}'_{\h}=\hat{p}'_{\h}-\alpha\,m\dot{X}$, $\dot{\hat{p}}'_{\h}=(\alpha-1)m\ddot{X}$, and $\ddot{\hat{x}}'_{\h}=-\ddot{X}$. Interestingly, the acceleration does not depend on $\alpha$, which means that the fictitious force $-m\ddot{X}$ can be viewed as deriving either from a {\em vector potential} (if $\alpha=1$), as in the formulation \eqref{SE'}, or from a {\em gravitational field} (if $\alpha=0$), as in \eqref{SE_relative}, or even from both elements simultaneously. The Hamiltonian \eqref{generalH'} is just an example of a gaugelike formulation which, as such, cannot be regarded as the ultimate physical representation of the dynamics. The object $\bra{\psi(0)}\ddot{\hat{x}}'_{\h}\ket{\psi(0)}$, on the other hand, seems to better incorporate the status of {\em law of motion}. Now, by adhering to this conceptual framework, we face no problem either with the Hamiltonian in Eq.~\eqref{SE'} or with its {\em mass-independent} generating transformation $\hat{\mathcal{G}}=e^{-iX\hat{p}/\hbar}$. Clearly, then, no issue can arise for the mass. One should also note that within the classical Hamiltonian formalism, we do have precisely the same transformations discussed above and an entirely equivalent mathematical structure. Accordingly, we could say that the Hamiltonian mechanics does not satisfies GC. Yet, we do not charge classical physics with any incompatibility with both {\em Galilean relativity} (GR) and EEP. This is so because the figure of merit is, in fact, the acceleration.

A second important point concerns the physical framework that is usually invoked to justify the existence of the residual phase. It has been suggested that this phase is a reminiscence of the twin-paradox effect in a nonrelativistic regime~\cite{greenberger01,coronado12}. The subtlety here, however, is that the resolution of the twin paradox is given in terms of accelerations of the reference frame, a physical phenomenon that necessarily demands interactions and an inertial reference frame in the background. Now, from a quantum mechanical viewpoint, physical interactions may generate correlations among the interacting systems. This process is fundamentally different from a mere change in the theoretical description and can, actually, preclude the existence of mass superpositions. Correlations also appear in the decay of unstable particles (see, e.g., Refs.~\cite{nystrand00,eberly05,ruza10}), which is a phenomenon that is taken as a primordial motivation for the existence of mass superpositions~\cite{greenberger01}. To better appreciate this point and its consequence, let us consider a system composed of an excited atom of mass $M$ and no radiation. In terms of energy eigenstates, the global state is $\ket{\psi_0}=\ket{M\!c^2}\ket{0}$. Before any measurement and at times comparable to the emission mean life, the system evolves to $\ket{\psi_t}=a\ket{M\!c^2}\ket{0}+be^{i\theta}\ket{mc^2}\ket{\hbar\omega}$, where $\omega$ is the excitation frequency, $b^2=1-a^2$ is the probability of emission, $\theta$ is an arbitrary phase, and $M\!c^2\cong mc^2 +\hbar\omega$ reflects the energy-mass conservation (the kinetic energy of the atom has been neglected). Now, this entangled state is rather different from the usually presumed mass superposition $a\ket{M\!c^2}+be^{i\theta}\ket{mc^2}$. In fact, for a state such as $\ket{\psi_t}$, which is typical of any decay process, it is impossible to access the phase $\theta$ in any interferometric experiment involving only the atom, as its reduced state is a statistical mixture. This means that we can say at best that in each run of the experiment, the atom has mass of {\em either} $M$ {\em or} $m$; it is not in a genuine superposition of both masses {\em in the same run}. We see, therefore, that even the justification for the existence of a mass superposition is debatable in the first place.

%-------------------------------------------
\section{Light reference frames \label{LRF}}
\vspace{-0.2cm}
Now we want to push the issue one step further. Considering abstract coordinate systems whose motion is immutable by principle, as if these reference systems were not themselves susceptible to physical interactions, is a good approximation in uncountable situations, particularly in regimes in which the reference frame is much heavier than the system. However, this is by no means the most fundamental approach one can admit, especially where the quantum domain is concerned. Here we want to consider as reference frames finite-mass bodies initially prepared in some quantum state relatively to a {\em primordial reference system} $\mathbb{S}_0$~\footnote{Throughout this paper, $\mathbb{S}_0$ is considered an ideally infinite-mass laboratory within which the state of the system is prepared. It could be alternatively regarded as an auxiliary {\em absolute frame of reference} to be posteriorly abandoned~\cite{aharonov84,angelo11,angelo12}.}. We then ask how the physics looks from the perspective of a quantum particle (quantum reference frame) freely moving relatively to another one. This scenario defines what we call a {\em quantum Galilei boost}.

The notion of quantum reference frames was introduced by Aharonov and Kaufherr~\cite{aharonov84}, who showed that it is possible to consistently formulate quantum theory without appealing to classical reference systems. Later on, it was shown that superselection rules commonly derive from the lack of a quantum reference frame (see~\cite{spekkens07} and references therein). More recently, the subject was revisited by one of us and collaborators~\cite{angelo11,angelo12} and fundamental contributions were provided in the field of quantum correlations and foundations of quantum mechanics. Here we investigate the compatibility of the laws of quantum mechanics with relativity principles upon quantum Galilei boosts.

We consider coordinate transformations in a scenario involving the following participants: a particle $\mathbb{P}$ of mass $m$, a system $\mathbb{S}$ of mass $M$, to be eventually promoted to a reference frame, and $\mathbb{S}_0$. Again, $\mathbb{S}_0$ is the primordial inertial reference frame, a classical-like body with ideally infinite mass, within which the quantum state of the composite system $\mathbb{S}+\mathbb{P}$ is prepared. From the perspective of $\mathbb{S}_0$, the Hamiltonian reads
\be
\hat{H}=\frac{\hat{P}^2}{2M}+V(\hat{X})+\frac{\hat{p}^2}{2m},
\label{HS0}
\ee
where $\hat{V}=V(\hat{X})$ is some potential to which $\mathbb{S}$ is submitted. To obtain the description relative to $\mathbb{S}$, we employ the mass-independent unitary generator introduced by Aharonov and Kaufherr in Ref.~\cite{aharonov84}, $\hat{\mathcal{G}}_{AK}=e^{-i\hat{X}\hat{p}/\hbar}$. The resulting coordinates are
\be\begin{array}{lcl}
\hat{X}'=\hat{X}, & \qquad & \hat{P}'=\hat{P}+\hat{p},\\ \\
\hat{x}'=\hat{x}-\hat{X}, & & \hat{p}'=\hat{p}.
\end{array}\label{x'AK}\ee
The Hamiltonian in the passive picture is
\be
\hat{H}'=\frac{\hat{P}'^2}{2M}+V(\hat{X}')+\frac{\hat{p}'^2}{2\mu}-\frac{\hat{P}'\hat{p}'}{M},
\label{H'AK}
\ee
where $\mu=mM/M_T$ and $M_T=M+m$. In this case, the Hamiltonian is not given in terms of the wanted coordinates. In fact, while $\hat{x}'$ is the position of $\mathbb{P}$ relative to $\mathbb{S}$, $\hat{p}'$ is the momentum of $\mathbb{P}$ relative to $\mathbb{S}_0$. In addition, in virtue of the coupling $\hat{P}'\hat{p}'$, it is impossible to elect a term that would separately account for the relative physics, even if $\hat{V}=0$. Clearly, GC is not respected. Also, as far as EEP is concerned, there is no sign of a gravitational field in the Hamiltonian \eqref{H'AK}. However, these conclusions change when we look at the acceleration of the particle. Indeed, from $\dot{\hat{x}}'_{\h}=\frac{\hat{p}'_{\h}}{\mu}-\frac{\hat{P}'_{\h}}{M}$, $\dot{\hat{p}}'_{\h}=0$, $\dot{\hat{X}}'_{\h}=\frac{\hat{P}'_{\h}}{M}-\frac{\hat{p}'_{\h}}{M}$, and $\dot{\hat{P}}_{\h}=-\partial_{\hat{X}'_{\h}}V(\hat{X}'_{\h})$, one shows that 
\be
\ddot{\hat{x}}'_{\h}=\tfrac{1}{M}\partial_{\hat{X}'_{\h}}V(\hat{X}'_{\h})=\tfrac{1}{M}\partial_{\hat{X}_{\h}}V(\hat{X}_{\h})=-\ddot{\hat{X}}_{\h},\nonumber
\ee
which is an expression of EEP. If $\hat{V}=0$, GC is retrieved, as $\ddot{\hat{x}}_{\h}=\ddot{\hat{x}}'_{\h}=0$. Moreover, the compatibility with the relativity principles is ensured for any initial state.

Of course, the above transformation is just one among many alternatives. Another unitary generator is~\cite{angelo12}
\be
\hat{\mathcal{G}}=\exp\left(-\frac{i\hat{X}\hat{p}}{\hbar}\right)\exp\left(\frac{i}{\hbar}\frac{m}{M_T}\hat{P}\hat{x}\right),
\label{G}
\ee
which has a clear connection with \eqref{GX}. This generator gives the well-known transformation
\be\begin{array}{lll}
\hat{X}'=\frac{M\hat{X}+m\hat{x}}{M+m}, &\qquad & \hat{P}'=\hat{P}+\hat{p}, \\ \\
\hat{x}'=\hat{x}-\hat{X}, & & \hat{p}'=\mu\left(\frac{\hat{p}}{m}-\frac{\hat{P}}{M}\right),
\end{array}\label{x'R}\ee
where $(\hat{x}',\hat{p}')$ are now the correct coordinates of $\mathbb{P}$ relative to $\mathbb{S}$.
The Hamiltonian in the passive picture assumes the form 
\be
\hat{H}'=\frac{\hat{P}'^2}{2M_T}+V\big(\hat{X}'-\tfrac{m}{M_T}\hat{x}'\big)+\frac{\hat{p}'^2}{2\mu}.
\label{H'2part}
\ee 
Once again, $\ddot{\hat{x}}'_{\h}=-\ddot{\hat{X}}_{\h}$. The benefit deriving from the generator \eqref{G} is evident: the transformed Hamiltonian is written in terms of coordinates accessible from the frame $\mathbb{S}$, so that we can still look for a consistent relative Hamiltonian formulation. To this end, let us take $\hat{V}=0$. It becomes clear that GC can be claimed only if one admits the replacement of masses $(M,m)\to (M_T,\mu)$. But this is a reasonable requirement, once it is well known from the two-body problem that the relative physics depends on the reduced mass. In fact, this requirement is unnecessary only when the reference frame is much heavier than the particle ($\tfrac{m}{M}\to 0$), a regime considered in all previous works on this subject. Concerning EEP, assume that $V(\hat{X})=-Mg\hat{X}$, so that $\mathbb{S}$ is moving with constant acceleration $-g$ relative to $\mathbb{S}_0$. In this case, the Hamiltonian \eqref{H'2part} becomes
\be
\hat{H}'=\frac{\hat{P}'^2}{2M_T}-Mg\hat{X}'+\frac{\hat{p}'^2}{2\mu}+\mu g\hat{x}'.\nonumber
\ee 
The appearance of the gravitational energy $\mu g \hat{x}'$ in $\mathbb{S}$'s perspective signals the compatibility of the Schr\"odinger equation with EEP. Note that the relation $\ddot{\hat{x}}'_{\h}=-\ddot{\hat{X}}_{\h}$ implies compatibility even when the frame $\mathbb{S}$ possesses a time-dependent acceleration in relation to $\mathbb{S}_0$, with this being, therefore, a more general statement.

%------------------------------------------
\section{Quantum Galilei boost \label{QGB}}
\vspace{-0.2cm}
We now examine a Galilei boost between quantum reference frames. To this end, we consider three physical systems, $\mathbb{S}_1$, $\mathbb{S}_2$, and $\mathbb{P}$, with respective masses $M_1$, $M_2$, and $m$. The first two systems will eventually play the role of reference frames, whose descriptions we want to compare. Once again, we elect a primordial reference frame $\mathbb{S}_0$ relative to which the following Hamiltonian holds:
\be
\hat{H}=\frac{\hat{P}_1^2}{2M_1}+V(\hat{X}_2-\hat{X}_1)+\frac{\hat{P}_2^2}{2M_2}+\frac{\hat{p}^2}{2m}.
\label{HSS'}
\ee
Here, however, there is no interaction between $\mathbb{S}_0$ and the system, so that the former can eventually be abandoned. Consider the following two sets of canonically conjugated operators:
\be\begin{array}{lll}
\hat{X}_1'=\frac{m\hat{x}+M_1\hat{X}_1+M_2\hat{X}_2}{M_T}, & & \hat{P}_1'=\hat{p}+\hat{P}_1+\hat{P}_2,\\ \\
\hat{X}_2'=\hat{X}_2-\hat{X}_1,& & \hat{P}_2'=M_2\left(\frac{\hat{P}_2}{M_2}-\frac{\hat{P}_1'}{M_T}\right)\!,\\ \\
\hat{x}'=\hat{x}-\hat{X}_1,& & \hat{p}'=m\left(\frac{\hat{p}}{m}-\frac{\hat{P}_1'}{M_T}\right),
\end{array}\qquad
\label{set1}
\ee
and
\be\begin{array}{lll}
\hat{X}_1'=\frac{m\hat{x}+M_1\hat{X}_1+M_2\hat{X}_2}{M_T}, & & \hat{P}_1'=\hat{p}+\hat{P}_1+\hat{P}_2,\\ \\
\hat{X}_2'=\frac{M_2}{\mu_2}\left(\hat{X}_2-\hat{X}_1'\right),& & \hat{P}_2'=\mu_2\left(\frac{\hat{P}_2}{M_2}-\frac{\hat{P}_1}{M_1}\right)\!,\\ \\
\hat{x}'=\frac{m}{\mu}\left(\hat{x}-\hat{X}_1'\right),& & \hat{p}'=\mu\left(\frac{\hat{p}}{m}-\frac{\hat{P}_1}{M_1}\right),
\end{array}\qquad
\label{set2}
\ee
where $M_T=m+M_1+M_2$, $\mu=\frac{mM_1}{m+M_1}$, and $\mu_2=\frac{M_2M_1}{M_2+M_1}$ are reduced masses, and $(\hat{X}_1',\hat{P}_1')$ turn out to be the center-of-mass observables relative to $\mathbb{S}_0$ in both sets. The respective generators of the transformations \eqref{set1} and \eqref{set2} are
\be
\hat{\mathcal{G}}_{1}(\mathbb{S}_1)=e^{-i\left(\frac{M_2}{M_T}\hat{X}_2+\frac{m}{M_T}\hat{x}\right)\frac{\hat{P}_1}{\hbar}}e^{i\frac{\hat{X}_1}{\hbar}(\hat{P}_2+\hat{p})}
\label{G1}
\ee
and
\be
&&\hat{\mathcal{G}}_{2}(\mathbb{S}_1)=e^{i\hat{\Lambda}}e^{\frac{i M_1}{M_T}\frac{\hat{X}_1\left(\hat{P}_2+\hat{p}\right)}{\hbar}}e^{-\left(\frac{M_2}{M_1}\hat{X}_2+\frac{m}{M_1}\hat{x}\right)\frac{i\hat{P}_1}{\hbar}},\label{G2}\\
&&\hat{\Lambda}=\frac{\lambda}{2\hbar}\{\hat{x},\hat{p}\} +\sum_{k=1,2} \frac{\lambda_k}{2\hbar}\{\hat{X}_k,\hat{P}_k \},\nonumber
\ee
where $\lambda=\ln\left(\frac{\mu}{m}\right)$, $\lambda_1=\ln\left(\frac{M_T}{M_1}\right)$, $\lambda_2=\ln\left(\frac{\mu_2}{M_2}\right)$, and $\{\hat{X}_k,\hat{P}_k\}=\hat{X}_k\hat{P}_k+\hat{P}_k\hat{X}_k$. We constructed these unitary operators based on the fact that momentum is the generator of displacements. The operator $\hat{\Lambda}$ appears in the second generator because the Jacobian of the transformation $\mathbf{r}\to\mathbf{r}'$ is not the unit.

Now, a fundamental point emerges. The physical coordinates, i.e., the ones accessible from $\mathbb{S}_1$, are $(\hat{X}_2',\hat{x}')$ of set \eqref{set1} and $(\hat{P}_2',\hat{p}')$ of set \eqref{set2}. However, they are not canonically conjugated to each other, as $[\hat{X}_2',\hat{p}']=i\hbar\tfrac{m}{M_1+m}$ and $[\hat{x}',\hat{P}_2']=i\hbar\tfrac{M_2}{M_1+M_2}$. This effect, discovered in Ref.~\cite{angelo11}, derives from the lightness of the reference frame and introduces an intrinsic inseparability in the joint Hilbert space. (In fact, this effect disappears when $\mathbb{S}_1$ is sufficiently heavy.) It follows, therefore, that for general quantum systems, with more than two particles, we should have no hope to obtain a Hamiltonian operator that is given in terms only of the coordinates accessible to the quantum frame of reference. In this sense, covariance will not be observed in the Hamiltonian description. Indeed, using the sets \eqref{set1} and \eqref{set2}, we obtain
\be
\hat{H}'=\frac{\hat{P}_1'^2}{2M_T}+V(\hat{X}_2')+\frac{\hat{P}_2'^2}{2\mu_2}+\frac{\hat{p}'^2}{2\mu}+\frac{\hat{P}_2'\hat{p}'}{M_1}
\label{H'1}
\ee
and
\be
\hat{H}'=\frac{\hat{P}_1'^2}{2M_T}+V\left(\hat{X}_2'+\tfrac{\mu}{M_1}\hat{x}'\right)+\gamma\left(\frac{\hat{P}_2'^2}{2\mu_2}+\frac{\hat{p}'^2}{2\mu}-\frac{\hat{P}_2'\hat{p}'}{M_1}\right), \nonumber \\
\label{H'2}
\ee
respectively, where $\gamma=\frac{M_1M_2m}{M_T\mu_2\mu}$. In both cases, the center-of-mass kinetic energy decouples. Still, none of these Hamiltonian operators is written only in terms of the coordinates relative to $\mathbb{S}_1$. Once again, however, we should observe that despite the manifest noncovariance of Schr\"odinger's equation for finite-mass references frames, we cannot maintain that quantum mechanics is not consistent with GC and EEP, for the Heisenberg equations give
\be
\ddot{\hat{x}}'_{\h}=-\frac{1}{M_1}\partial_{\hat{X}_{2_{\h}}'}V(\hat{X}_{2_{\h}}')=-\ddot{\hat{X}}_{1_{\h}},
\ee
for \eqref{H'1}, and
\be
\frac{\dot{\hat{p}}'_{\h}}{\mu}=-\ddot{\hat{X}}_{1_{\h}},
\ee
for \eqref{H'2}. Note that in the latter case, the pertinent object is the time derivative of momentum because it is the relative quantity [see set \eqref{set2}]. Again, it is clear that the relations $\bra{\psi(0)}\ddot{x}'_{\h}\ket{\psi(0)}$ and $\bra{\psi(0)}\dot{p}'_{\h}/\mu\ket{\psi(0)}$ are in total harmony with GC and EEP, even within the context of quantum reference frames with arbitrary masses. 

So far we have discussed the physics seen by the finite-mass reference system $\mathbb{S}_1$. Now we are ready to make the boost to $\mathbb{S}_2$'s view. Since we have employed no special assumption on the masses involved, it is trivial to infer that the same conclusions we arrived at above will hold from the perspective of the other finite-mass reference system $\mathbb{S}_2$. Formally,  the way to move to the physics seen by $\mathbb{S}_2$ is via the generator $\hat{\mathcal{G}}_{k'}(\mathbb{S}_2)\,\hat{\mathcal{G}}_k^{\dag}(\mathbb{S}_1)$ $(k,k'=1,2)$. This will give the physics of the center of mass of the global system $\mathbb{S}_1+\mathbb{S}_2+\mathbb{P}$ relative to $\mathbb{S}_0$ along with the physics of $\mathbb{S}_1$ and $\mathbb{P}$ relative to $\mathbb{S}_2$. Alternatively, we can explore the following conceptually different strategy for the implementation of the boost. 

Assume that $\mathbb{S}_0$ is an {\em absolute reference frame}~\cite{aharonov84,angelo11,angelo12}, i.e., an immaterial coordinate system. As such, it plays the auxiliary role of providing an ideally inertial description typical of infinite-mass frames. In this case, the state of the system can be thought of as having been prepared by some apparatus, no longer present, or, e.g., by $\mathbb{S}_1$~\footnote{If prepared by $\mathbb{S}_1$, the state will be in the form $\ket{\psi_{cm}}_{\mathbb{S}_0}\otimes\ket{\psi_{\mathbb{S}_2\mathbb{P}}}_{\mathbb{S}_1}$, as the center of mass cannot be prepared by any interaction internal to the system. Moving to the absolute coordinates one would have $\ket{\psi_{\mathbb{S}_1\mathbb{S}_2\mathbb{P}}}_{\mathbb{S}_0}$, an entangled state in general.}.  The {\em real} physics is then given by the Hamiltonians \eqref{H'1} and \eqref{H'2}, provided we neglect the absolute center-of-mass energy $\hat{P}_1'^2/2M_T$. Thus, hereafter we focus, for simplicity, on the relevant part of the Hamiltonian \eqref{H'1},
\be
\hat{H}_{\mathbb{S}_1}'=\frac{\hat{P}_2'^2}{2\mu_2}+V(\hat{X}_2')+\frac{\hat{p}'^2}{2\mu}+\frac{\hat{P}_2'\hat{p}'}{M_1}.
\ee 
Now, one can apply the generator \eqref{G} with the appropriate replacements: $(\hat{X},\hat{P})\to(\hat{X}_2',\hat{P}_2')$ and $M_T\to M_2+m$. In this case, the resulting description will be composed of the center of mass of the system $\mathbb{S}_2+\mathbb{P}$ relative to $\mathbb{S}_1$, plus the physics of $\mathbb{P}$ relative to $\mathbb{S}_2$. The transformation reads
\be\begin{array}{lll}
\hat{X}''_2=\frac{m \hat{x}'+M_2\hat{X}'_2}{m+M_2}, & & \hat{P}''_2=\hat{p}'+\hat{P}', \\ \\
\hat{x}''=\hat{x}'-\hat{X}'_2, & & \hat{p}''=\mu'\left(\frac{\hat{p}'}{m}-\frac{\hat{P}'_2}{M_2} \right),
\end{array}\ee
where $\mu'=\tfrac{mM_2}{m+M_2}$. In terms of the absolute coordinates, one has the relations $\hat{x}''=\hat{x}-\hat{X}_2$ and $\hat{p}''=\mu'\left(\frac{\hat{p}}{m}-\frac{\hat{P}_2}{M_2} \right)$, which show that $(\hat{x}'',\hat{p}'')$ indeed correspond to coordinates relative to $\mathbb{S}_2$. The Hamiltonian transforms as
\be
\hat{H}''=\frac{\hat{P}_2''^2}{2\mu_2'}+V\left(\hat{X}_2''-\tfrac{m}{m+M_2}\hat{x}''\right)+\frac{\hat{p}''^2}{2\mu'},\quad
\ee
where $\mu_2'=\tfrac{M_1(m+M_2)}{M_1+(m+M_2)}$. From $\dot{\hat{x}}''_{\h}=\hat{p}''_{\h}/\mu'$ and $\dot{\hat{p}}''_{\h}=\tfrac{\mu'}{M_2}\partial_{\hat{X}'_{2_{\h}}}V_2(\hat{X}'_{2_{\h}})$, it is easy to show that $\ddot{\hat{x}}''_{\h}=-\ddot{\hat{X}}_{2_{\h}}$. This is the result expected for the motion of $\mathbb{P}$ relative to $\mathbb{S}_2$. In particular, if $\hat{V}=0$, it follows from the present analysis that 
\be
\bra{\psi(0)}\ddot{x}''_{\h}\ket{\psi(0)}=\bra{\psi(0)}\ddot{x}'_{\h}\ket{\psi(0)}=0.
\ee
This relation is the proof that the physics generated by quantum mechanics for the motion of a particle is indeed {\em covariant} under Galilei boosts between two distinct quantum reference frames.

%---------------------------
\section{Concluding remarks}
\vspace{-0.2cm}
Schr\"odinger's equation inherited, from classical mechanics, a Hamiltonian structure with many benefits and some subtleties. Canonical transformations, in particular, are known to constitute an important mathematical tool that allows for explicit integration of the equations of motion in a number of relevant problems. However, the physical interpretation underlying the transformed structure is often not obvious. In addition, by either adding constant terms or adopting specific gauges, one can construct an infinite number of Hamiltonian functions for the same physical dynamics. In this paper, we show that the longstanding tension between quantum mechanics and both {\em Galilean covariance} (GC) and {\em Einstein's equivalence principle} (EEP) dissipates when submitted to a careful analysis of the quantum mechanical structure and its connection with observable quantities. By relying on standard concepts of quantum theory, such as unitary generators, quantum reference frames, and expectation values, we were able to identify the defining features of the aforementioned issue and solve it in a natural and intuitive way.

Specifically, our results can be summarized as follows. First, we identified a proper interpretation for unitary transformations employed within the vector state formalism. In particular, we distinguished between the passive and active pictures, which are often used simultaneously in a very inaccurate way. Second, within this framework, we showed that the Schr\"odinger equation is fully compatible with GC and EEP when we consider Galilei boosts between infinite-mass reference frames. Also, we argued that Bargmann's superselection rule is not demanded by the formalism, provided we stick to the standard notion of mass as a $c$ number. Moreover, by explicitly exhibiting an example in which an invariant dynamics emerges from a gaugelike Hamiltonian formulation, which interpolates between vector potentials and gravitational fields, we showed that the experimentally accessible expectation value $\bra{\psi(0)}\ddot{\hat{x}}'\ket{\psi(0)}$ better encompasses the physical features of a quantum law of motion. Third, we defined the notion of a {\em quantum Galilei boost}, i.e., a transformation of coordinates that brings the physical description to the perspective of a finite-mass reference frame, which is itself describable by the quantum theory. This fundamental approach remarkably exposes the limitations of the Hamiltonian formulation in respecting GC, while the Schr\"odinger equation and the respective equations of motion reveal full compatibility with {\em Galilean relativity} (GR) and EEP. In addition, our analysis shows that quantum mechanics does not need to rely on classical-like notions, such as infinite-mass reference frames, to correctly describe the relativity of motion in the microscopic realm.

Finally, it is worth emphasizing that we should not regard the sophisticated form of the transformed Hamiltonian operators as a symptom of violation of any physical principle because this is the way the Hamiltonian formalism implements the lightness of reference frames. Actually, it is immediately conclusive from our results, which are based exclusively on canonical transformations, that in classical mechanics, we have exactly the same manifestations of Hamiltonian noncovariance. However, no one is willing to deny the compatibility between classical mechanics and GR based on this fact. Most probably, rather than concluding that classical physics violates GC, many would say, along with us, that it is the Hamiltonian formalism that does so.

\acknowledgments
\vspace{-0.2cm}
S.T.P. acknowledges support of a scholarship from UFPR-TN and R.M.A. acknowledges financial support from the National Institute for Science and Technology of Quantum Information (INCT-IQ; CNPq/Brazil). Discussions with F. P. Devecchi are gratefully acknowledged.

%------------------------------------------------------------------------

%-------------------------------------------------------

\begin{thebibliography}{99}
\bibitem{bargmann54} U. Bargmann, Ann. Math. {\bf 59}, 1 (1954).
\bibitem{ballentine98} L. E. Ballentine, {\em Quantum Mechanics: A Modern Development} (World Scientific, Singapore, 1998).
\bibitem{yan03} K. Gottfried and T.-M. Yan, {\em Quantum Mechanics: Fundamentals} (Springer, New York, 2003).
\bibitem{greenberger01} D. M. Greenberger, Phys. Rev. Lett. {\bf 87}, 100405 (2001).
\bibitem{okon13} H. Hernandez-Coronado and E. Okon, Phys. Lett. A {\bf 377}, 2293 (2013).
\bibitem{leblond63} J.-M. Levy-Leblond, J. Math. Phys. {\bf 4}, 776 (1963).
\bibitem{greenberger70_1} D. M. Greenberger, J. Math. Phys. {\bf 11}, 2329 (1970).
\bibitem{greenberger70_2} D. M. Greenberger, J. Math. Phys. {\bf 11}, 2341 (1970).
\bibitem{greenberger74_1} D. M. Greenberger, J. Math. Phys. {\bf 15}, 395 (1974).
\bibitem{greenberger74_2} D. M. Greenberger, J. Math. Phys. {\bf 15}, 406 (1974).
\bibitem{pascual90} A. Galindo and P. Pascual, {\em Quantum Mechanics I} (Springer-Verlag, Berlin, 1990).
\bibitem{giulini96} D. Giulini, Ann. Phys. {\bf 249}, 222 (1996).
\bibitem{holland99} H. R. Brown and P. R. Holland, Am. J. Phys. {\bf 67}, 204 (1999).
\bibitem{brown99} H. R. Brown, in {\em From Physics to Philosophy}, edited by J. Butterfield and C. Pagonis (Cambridge University Press, Cambridge, 1999) pp. 45--70.
\bibitem{wick07} N. L. Harshman and S. Wickramasekara, Phys. Rev. Lett. {\bf 98}, 080406 (2007).
\bibitem{pad11} H. Padmanabhan and T. Padmanabhan, Phys. Rev. D {\bf 84}, 085018 (2011).
\bibitem{coronado12} H. Hernandez-Coronado, Found. Phys. {\bf 42}, 1350 (2012).
\bibitem{unnik02} C. S. Unnikrishnan, Mod. Phys. Lett. A {\bf 17}, 1081 (2002).
\bibitem{sudarsky03} C. Chryssomalakos and D. Sudarsky, Gen. Rel. Grav. {\bf 35}, 605 (2003).
\bibitem{wawr11} A. Herdegen and J. Wawrzycki, Phys. Rev. D {\bf 66}, 044007 (2002).
\bibitem{aharonov84} Y. Aharonov and T. Kaufherr, Phys. Rev. D {\bf 30}, 368 (1984).
\bibitem{spekkens07} S. D. Bartlett, T. Rudolph, and R. W. Spekkens, Rev. Mod. Phys. {\bf 79}, 555 (2007).
\bibitem{angelo11} R. M. Angelo, N. Brunner, S. Popescu, A. J. Short, and P. Skrzypczyk, J. Phys. A {\bf 44}, 145304 (2011).
\bibitem{angelo12} R. M. Angelo and A. D. Ribeiro, J. Phys. A {\bf 45}, 465306 (2012).
\bibitem{nystrand00} S. R. Klein and J. Nystrand, Phys. Rev. Lett. {\bf 84}, 2330 (2000).
\bibitem{eberly05} M. V. Fedorov, M. A. Efremov, A. E. Kazakov, K. W. Chan, C. K. Law, and J. H. Eberly, Laser Phys. {\bf 15}, 1229 (2005).
\bibitem{ruza10} J. Ruz\v{a}, Physica E {\bf 42}, 327 (2010).
\end{thebibliography}
\end{document}